\documentstyle[12pt]{article}
\input epsfig
\begin{document}
\date{}

\author{V. Yu. Petrov and P. V. Pobylitsa  \\
{ \small\em Petersburg Nuclear Physics Institute, Gatchina,
188350, Russia}
}
\title{Pion wave function from the instanton vacuum
model}
\maketitle

\begin{abstract}

The pion wave function is computed in the low-energy effective
theory inspired by the instanton vacuum model. The result is numerically
rather close to the asymptotic wave function.

\end{abstract}

For a long time the most popular models for the pion wave function were the
asymptotic wave function
corresponding to the asymptotically large normalization
points and the wave function suggested by Chernyak-Zhitnitsky~\cite{CZ}
which is a
superposition of two Gegenbauer polynomials. The recent CLEO
experiment~\cite{CLEO,Savinov}
seems to exclude the wave function of Chernyak-Zhitnitsky. Although the
asymptotic wave function is compatible with the phenomenological data, it is
hard to believe that the asymptotic limit is already reached in the current
experiments with rather low $Q^2$.

In this letter we compute the pion wave function in the low-energy effective
theory~\cite{DP-quarks-NP} inspired by the instanton vacuum
model~\cite{DP-pure-gluonic}. Although the underlying
low-energy effective theory has nothing to do with the large $Q^2$
asymptotic limit of QCD our result is numerically rather close to the
asymptotic wave function.

In QCD the pion wave function $\phi (\zeta )$ is given by
\[
\phi (\zeta )=\frac 1{i\sqrt{2}F_\pi }\int\limits_{-\infty }^\infty \frac{
d\tau }{2\pi }e^{-i\tau \zeta (nP)}
\]
\begin{equation}
\times \langle 0|\bar d(\tau n)\,n_\mu \gamma ^\mu \gamma _5\exp \left\{
ig\int\limits_{-\tau }^\tau d\tau ^{\prime }\,n^\lambda A_\lambda (\tau
^{\prime }n)\right\} u(-\tau n)|\pi ^{+}(P)\rangle \quad \quad (n^2=0)
\,.
\label{phi-def-2}
\end{equation}
Here $n$ is a lightcone vector corresponding to the infinite momentum frame
where the quark and antiquark carry momentum fractions $(1\pm \zeta )/2$.
This wave function is normalized by the condition
\begin{equation}
\int\limits_{-1}^1d\zeta \phi (\zeta )=1
\,.
\label{phi-normalization}
\end{equation}

We want to compute the pion wave function $\phi (\zeta )$ in the model
describing the interaction of quark fields $\psi $ with the pions $\pi $
with the effective lagrangian
\begin{equation}
L=\bar \psi (i\gamma ^\mu \partial _\mu -MU^{\gamma _5})\psi
\end{equation}
where $U$ is the chiral pion field
\begin{equation}
U=\exp (i\pi ^a\tau ^a/F_\pi )\,,\quad \quad F_\pi =f_\pi /\sqrt{2}
\,.
\end{equation}
This lagrangian is derived from the instanton vacuum model of QCD. The
effective quark mass $M$ appearing due to the spontaneous breakdown of the
chiral symmetry is momentum dependent so that the mass term in the above
lagrangian should be understood in the following sense
\begin{equation}
\int d^4x\,\bar \psi MU^{\gamma _5}\psi \rightarrow \int \frac{d^4k_1}{(2\pi
)^4}\frac{d^4k_2}{(2\pi )^4}\bar \psi (k_1)\sqrt{M(k_1)}U^{\gamma
_5}(k_2-k_1)\sqrt{M(k_2)}\psi (k_2)
\,.
\end{equation}
Let us start from the simplified calculation neglecting the momentum
dependence of the quark mass. In the leading order of the large number of
quark colors we have in this model
\[
\langle 0|\bar d(z)\gamma ^\mu \gamma _5u(-z)|\pi ^{+}(P)\rangle
\]
\[
=\sqrt{2}\frac M{F_\pi }N_c\int \frac{d^4k}{(2\pi )^4}e^{izk}e^{iz(k-P)}
{\rm Sp}\,\left[ \gamma ^\mu \gamma _5\frac i{\not k-M}\,\gamma
_5\frac i{(\not k-\not P)-M}\right]
\]
\begin{equation}
=-\sqrt{2}\frac{M^2}{F_\pi }4N_c\int \frac{d^4k}{(2\pi )^4}e^{iz(k-2P)}\,
\frac{P^\mu }{(k^2-M^2)[(P-k)^2-M^2]}
\,.
\label{matr-element-1}
\end{equation}
Inserting this result into the QCD\ expression for the pion wave function
(\ref{phi-def-2}) we find
\begin{equation}
\phi (\zeta )=-i\frac{M^2}{F_\pi ^2}4N_c\int \frac{d^4k}{(2\pi )^4}\delta
\left[ \frac{n\cdot (P-2k)}{(n\cdot P)}-\zeta \right] \frac
1{(k^2-M^2+i0)[(P-k)^2-M^2+i0]}
\,.
\label{phi-model-res}
\end{equation}
This result is derived for massless pions $P^2=0$, $n$ is also a light cone
vector $n^2=0$. Therefore the right hand side of (\ref{phi-model-res}) can
depend on $n$ and $P$ (at fixed $\zeta $) only through the Lorentz invariant
scalar $(n\cdot P)$. But the is right hand side of (\ref{phi-model-res})
obviously does not change under the dilatations $n\rightarrow \alpha n$.
Hence the rhs cannot depend on $(n\cdot P)$. Thus we conclude that as long
as $n$ and $P$ are light cone vectors the rhs of (\ref{phi-model-res}) does
not depend on their specific choice (of course we assume that $(n\cdot P)\ne
0$).

Taking into account that in our effective chiral model the pion decay
constant is given by the (euclidean) integral
\begin{equation}
F_\pi ^2=4N_cM^2\int \frac{d^4k}{(2\pi )^4}\frac 1{(k^2+M^2)^2}\Biggr|_{{\
{\rm Euclid}}}
\end{equation}
we immediately see that our result for the pion wave function (\ref
{phi-model-res}) obeys the normalization condition (\ref{phi-normalization}).

Using the light cone vectors $n$ and $P$ as a basis we introduce the
light-cone coordinates
\begin{equation}
k^{+}=(nX),\quad \quad k^{-}=2(PX)/(nP)
\end{equation}
so that (\ref{phi-model-res}) can be rewritten in the form

\[
\phi (\zeta )=-i\frac{M^2}{F_\pi ^2}2N_c\int \frac{dk^{+}dk^{-}d^2k^{\perp }
}{(2\pi )^4}\delta \left[ 1-\frac{2k^{+}}{P^{+}}-\zeta \right]
\]
\begin{equation}
\times \frac 1{(k^{+}k^{-}-|k^{\perp
}|^2-M^2+i0)[-P^{+}k^{-}+k^{+}k^{-}-|k^{\perp }|^2-M^2+i0]}
\,.
\end{equation}
We taking the integral over $k^{+}$ using the delta function
\[
\phi (\zeta )=-i\frac{M^2}{F_\pi ^2}N_cP^{+}\int \frac{dk^{-}d^2k^{\perp }}{
(2\pi )^4}
\]
\begin{equation}
\times \left( \frac{1-\zeta }2P^{+}k^{-}-|k^{\perp }|^2-M^2+i0\right)
^{-1}\left( -\frac{1+\zeta }2P^{+}k^{-}-|k^{\perp }|^2-M^2+i0\right) ^{-1}
\,.
\end{equation}
The integral over $k^{-}$ vanishes if both poles in $k^{-}$ lie in the same
complex half plane. This shows that our result for the pion wave function
differs from zero only in the interval $|\zeta |<1$ as it should be.
Performing the integral over $k^{-}$ we find
\begin{equation}
\phi (\zeta )=\frac{M^2}{F_\pi ^2}N_c\theta (1-\zeta ^2)\int \frac{
d^2k^{\perp }}{(2\pi )^3}\left( |k^{\perp }|^2+M^2\right) ^{-1}
\,.
\label{phi-res-1}
\end{equation}
Since we have already proved that our result for the pion wave function $
\phi (\zeta )$ obeys the normalization condition (\ref{phi-normalization})
we conclude that
\begin{equation}
\phi (\zeta )=\frac 12\theta (1-\zeta ^2)
\,.
\label{phi-step-function}
\end{equation}
Thus in the case of the momentum-independent effective quark mass the quark
wave function is simply a step function on the interval $|\zeta |<1$. This
result seems to be rather unphysical since the wave function does not vanish
at the edge points $\zeta =\pm 1$. Let us show the momentum dependent quark
mass $M(k)$ restores the physical behaviour at the edge points.

Although the instanton vacuum model of QCD leads to the momentum dependence
of the quark mass this dependence is rather slow and at small momenta one
can simply approximate $M(k)$ by its value at zero momentum $M(k)\approx
M(0) $. Therefore with good accuracy we can ignore the momentum dependence
of the mass in the denominators of the rhs of (\ref{phi-model-res}). On the
other hand we cannot ignore the momentum dependence in the numerator because
it is responsible both for the ultraviolet regularization of the integral
and for the vanishing of the pion amplitude at the edge points $\zeta =\pm 1$
. Therefore below we work with the following approximation
\[
\phi (\zeta )=-i\frac 1{F_\pi ^2}4N_c\int \frac{d^4k}{(2\pi )^4}\sqrt{
M(k)M(k-P)}\delta \left[ n\cdot (P-2k)-\zeta (n\cdot P)\right]
\]
\begin{equation}
\,\times \frac{(n\cdot k)M(k-P)+(n\cdot (P-k))M(k)}{
[k^2-M_0^2][(P-k)^2-M_0^2]}
\,.
\end{equation}
The momentum dependence of the effective quark mass computed in the
instanton vacuum model of QCD can be with good accuracy approximated by
\begin{equation}
M(k)=\frac{M_0}{(1+k^2\rho ^2)^2}
\,.
\label{M-model}
\end{equation}
Now we can generalize the above results to case of the momentum dependent
mass arriving at

Let us introduce compact notation
\begin{equation}
u\equiv \frac{1+\zeta }2|k_{\perp }|^2+\frac{1-\zeta }2|(P-k)_{\perp }|^2
\,.
\end{equation}
Then
\begin{equation}
\phi (\zeta )=\frac{(1-\zeta )(1+\zeta )}2\theta (1-|\zeta |)\frac{M_0^2\rho
^{-4}}{F_\pi ^2}N_c
\,,
\end{equation}
\begin{equation}
\times \int \frac{d^2k^{\perp }}{(2\pi )^3}\frac{u+\frac 12(M_0^2+\rho ^{-2})
}{\left( u+\rho ^{-2}\right) \left( u+\frac{1+\zeta }2\rho ^{-2}+\frac{
1-\zeta }2M_0^2\right) \left( u+M_0^2\right) \left( u+\frac{1+\zeta }2M_0^2+
\frac{1-\zeta }2\rho ^{-2}\right) }
\end{equation}
where we use the compact notation
\begin{equation}
u\equiv \frac{1+\zeta }2|k_{\perp }|^2+\frac{1-\zeta }2|(P-k)_{\perp }|^2
\,.
\end{equation}
It is easy to see that in the limit of the momentum independent mass  $\rho
\rightarrow 0$  we recover the old result (\ref{phi-step-function}). On the
other hand the momentum dependence of the quark mass leads to the vanishing
of the pion wave function at the edge points $\zeta =\pm 1$.

The result is presented at Fig. 1. We see that our model calculation is
rather close to the asymptotic wave function.

The measurements of the CLEO group allow to extract information on the
quantity
\begin{equation}
I_0 = 2 \int\limits_{-1}^{1} \frac{d\zeta}{1-\zeta^2} \phi_\pi(\zeta)
\,.
\end{equation}
The asymptotic wave function and the ansatz of
Chernyak--Zhitnitsky give respectively $I_0^{as}=3$ and $I_0^{CZ}=5$.
Our result
\begin{equation}
I_0=3.16
\end{equation}
is close to $I_0^{as}$.

The asymptotic wave function with the one-loop corrections
taken into account agrees \cite{MR}
 with the existent experimental data of CLEO
\cite{CLEO,Savinov,KR},
whereas the wave function of Chernyak--Zhitnitsky \cite{CZ} is
excluded.

Since our wave function happens to be close to the asymptotic there
is no need to repeat the calculations of the one-loop corrections,
so that the instanton wave function also agrees with the CLEO data.

\section*{Acknowledgements}

The authors are grateful to V.~Braun, D.~Diakonov, K.~Goeke,
N.~Kivel and M.~Polyakov for fruitful discussions.

The work has been supported in part by the grants INTAS-93-1630-EXT
and INTAS-RFBR-95-0681.

\begin{figure}
\centerline{\epsfig
  {figure=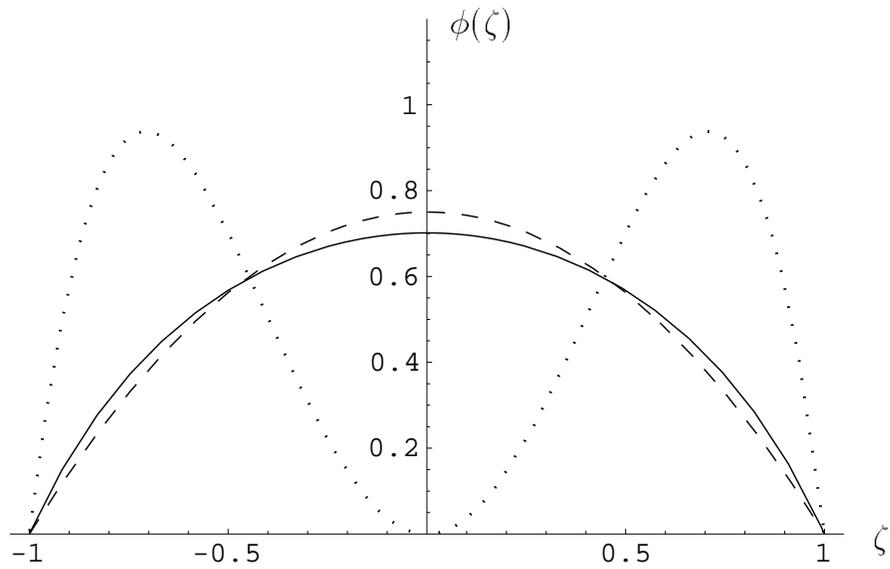,height=15cm,bbllx=0bp,bblly=400bp,bburx=550bp,bbury=792bp,clip=}}
\caption
{Pion wave function $\phi(\zeta)$ calculated in the instanton
vacuum model (solid) compared to the asymptotic (dashed) and
Chernyak--Zhitnitsky (dotted) wave functions.}
\end{figure}

\end{document}